\documentclass[10pt,aps,pre,twocolumn]{revtex4-1}

\usepackage{mathrsfs}
\usepackage{bbm}
\usepackage{amsthm}
\usepackage{amsmath}
\usepackage{amssymb}
\usepackage{graphicx}
\usepackage{mathtools}

\makeatletter

\usepackage{hyperref}
\usepackage{color}
\definecolor{myblue}{RGB}{0,50,200}
\hypersetup{
colorlinks,
citecolor=myblue,
linkcolor=myblue,
urlcolor=myblue
}
\allowdisplaybreaks

\newcommand{\mca}{\mathcal}
\newcommand{\mbb}{\mathbb}

\newcommand{\msf}{\mathsf}

\newcommand{\bra}[1]{\left( #1 \right)}
\newcommand{\bras}[1]{\left[ #1 \right]}

\newcommand{\pp}{\partial}

\newcommand{\Br}[1]{\langle #1|}
\newcommand{\Kt}[1]{|#1\rangle}
\newcommand{\Mat}[1]{|#1\rangle\langle #1|}
\newcommand{\BraKet}[2]{\langle #1|#2\rangle}
\newcommand{\MatBraKet}[3]{\langle #1|#2|#3\rangle}

\theoremstyle{definition}

\theoremstyle{remark}

\theoremstyle{result}
\newtheorem{result}{Result}

\begin{document}
\title{Toward relaxation asymmetry: Heating is faster than cooling}

\author{Tan Van Vu}
\email{tanvu@rk.phys.keio.ac.jp}
\altaffiliation{Current Address: Department of Physics, Keio University, 3-14-1 Hiyoshi, Kohoku-ku, Yokohama 223-8522, Japan}
\affiliation{Department of Information and Communication Engineering, Graduate
School of Information Science and Technology, The University of Tokyo,
Tokyo 113-8656, Japan}

\author{Yoshihiko Hasegawa}
\email{hasegawa@biom.t.u-tokyo.ac.jp}
\affiliation{Department of Information and Communication Engineering, Graduate
School of Information Science and Technology, The University of Tokyo,
Tokyo 113-8656, Japan}

\date{\today}

\begin{abstract}
An asymmetry in thermal relaxation toward equilibrium has been uncovered for Langevin systems near stable minima [Phys. Rev. Lett. 125, 110602 (2020)].
It has been shown that, given the same degree of nonequilibrium of the initial distributions, relaxation from a lower temperature state (heating) is faster than that from a higher temperature state (cooling).
In this study, we elucidate this relaxation asymmetry for discrete-state Markovian systems described by the master equation.
We rigorously prove that heating is faster than cooling for arbitrary two-state systems, whereas for systems with more than two distinct energy levels, the relaxation asymmetry is no longer universal.
Furthermore, for systems whose energy levels degenerate into two energy states, we find that there exist critical thresholds of the energy gap.
Depending on the magnitude of the energy gap, heating can be faster or slower than cooling, irrespective of the transition rates between states.
Our results clarify the relaxation asymmetry for discrete-state systems and reveal several hidden features inherent in thermal relaxation.
\end{abstract}

\pacs{}
\maketitle

\section{Introduction}
Systems attached to thermal reservoirs will relax toward a stationary state.
Such thermal relaxation processes are ubiquitous in nature and possess rich properties from both dynamic and thermodynamic perspectives.
One of the counterintuitive behaviors is the Mpemba effect \cite{Mpemba.1969.PE}, where cooling a hot system is faster than cooling a cold system.
Such nonmonotonic relaxation phenomena have been observed in various systems \cite{Auerbach.1995.AJP,Antonio.2017.PRL,BaityJesi.2019.PNAS,Kumar.2020.N,Biswas.2020.PRE} and theoretically analyzed for microscopic dynamics \cite{Lu.2017.PNAS,Klich.2019.PRX,Takada.2021.PRE}.
In addition, it was found that cooling a system before heating it could lead to exponentially fast relaxation \cite{Gal.2020.PRL}.
From the perspective of thermodynamics, thermal relaxation processes exhibit universal relations regarding irreversibility, which is quantified by irreversible entropy production \cite{Landi.2021.RMP}.
Notably, it has been shown that irreversible entropy production during thermal relaxation is lower-bounded by information-theoretical \cite{Alhambra.2017.PRA,Shiraishi.2019.PRL,Vu.2021.PRL2} and geometrical \cite{Vu.2021.PRL} distances between the initial and final states in both classical and quantum regimes.
These relations imply stronger inequalities than the conventional second law of thermodynamics and impose geometrical constraints on the possible relaxation path.
Since thermal relaxation is important in condensed matter \cite{Dattagupta.2012} and heat engines \cite{Benenti.2017.PR}, deepening our understanding of thermal relaxation would benefit research in these areas.

Consider preparing a thermal state corresponding to a given temperature via thermal relaxation; i.e., attaching a system to a single reservoir and allowing it to relax toward equilibrium.
In this setting, the relaxation time is a quantity of interest and can be approximately estimated via the convergence rate of the system state toward the equilibrium state \cite{Temme.2010.JMP}.
Given two identical systems initiated in thermal states, one at a lower and the other at a higher temperature than the given temperature, a natural question arises: which one relaxes faster?
Recently, this question has been addressed by Lapolla and Godec \cite{Lapolla.2020.PRL} for continuous-state Langevin systems.
By considering a pair of thermodynamically equidistant temperature quenches (which have the same nonequilibrium free energy difference), they unveiled an unforeseen asymmetry in thermal relaxation; i.e., relaxation from a lower temperature is faster than that from a higher temperature.
Roughly speaking, it implies that heating up cold objects is faster than cooling down hot objects.
This phenomenon has been proven for quenches of dynamics near stable minima; however, it is not universal for generic systems because counterexamples have been constructed using multi-well potentials \cite{Lapolla.2020.PRL}. Another recent study \cite{Meibohm.2021.PRE} has reported that heating can occur faster or slower than cooling even for anharmonic single-well potentials, and a crossover region emerges if the quenches are not too far from equilibrium.
Nonetheless, the relaxation asymmetry may be universal for a specific class of multi-well potentials. It is well known that overdamped diffusion under a multi-well potential with sufficiently high barriers converges to Markov jump dynamics at long times. By projecting the Langevin dynamics onto the Markov jump process between basins, it has been shown that the general asymmetry is preserved in degenerate potentials with separated time scales \cite{Lapolla.2020.PRL}.

Since the average energy of a thermal state increases with the temperature, the relaxation asymmetry allows us to say that, from the energetic perspective, uphill relaxation is faster than downhill relaxation, which is counterintuitive to an extent.
Moreover, relaxation speed cannot be characterized solely by thermodynamic quantities such as dissipation or frenesy \cite{Maes.2020.PR}.
Therefore, it is highly nontrivial that free energy plays an essential role as a quantifier of nonequilibrium degree in equidistant temperature quenches.

In this study, we elucidate the relaxation asymmetry for discrete-state systems modeled by Markov jump processes, thus improving our understanding of thermal relaxation.
First, we prove that heating is faster than cooling in an arbitrary two-state system, affirming that there is universality in the relaxation asymmetry.
However, we find that it is not the case for generic systems with at least three distinct energy levels.
By analytically constructing counterexamples, we demonstrate that heating can be faster or slower than cooling, depending on the transition rates.
Nevertheless, restricted to a particular class of systems, some universal results on the relaxation asymmetry are obtained.
We show that when the energy levels of the system are two-state degenerate, there exist two critical energy gap thresholds.
Depending on whether the energy gap is larger or smaller than these thresholds, it can be concluded with certainty that heating is faster or slower than cooling.
These theoretical results are numerically demonstrated using several discrete-state systems.

\section{Setup}
We consider the thermal relaxation process of an open system with $N$ states.
The system is coupled to a thermal reservoir at the inverse temperature $\beta_{f}=(k_{\rm B}T_{f})^{-1}$, where $k_{\rm B}$ is the Boltzmann constant.
Owing to interaction with the thermal reservoir, stochastic transitions between states are induced.
The dynamics of the system is governed by the master equation,
\begin{equation}
\pp_t\Kt{p_t}=\msf{R}\Kt{p_t},
\end{equation}
where $\Kt{p_t}\coloneqq [p_1(t),\dots,p_N(t)]^\top$ denotes the probability distribution of the system at time $t$; the matrix $\msf{R}=[R_{mn}]\in\mbb{R}^{N\times N}$ is time-independent with $R_{mn}\ge 0$ denoting the transition rate from state $n$ to state $m\,(\neq n)$, and $\sum_{m}R_{mn}=0$.
Without loss of generality, we assume that $E_1\ge\dots\ge E_N$, where $E_n$ denotes the energy of state $n$.
The transition rates satisfy the detailed balance condition, $R_{mn}e^{-\beta_{f} E_n}=R_{nm}e^{-\beta_{f} E_m}$, which is a sufficient condition such that the system always relaxes to the thermal Gibbs state $\Kt{\pi^{f}}$ after a sufficiently long time, irrespective of the initial state.
Here, 
\begin{equation}
\pi_n^{f}\coloneqq\frac{e^{-\beta_{f} E_n}}{Z_{\beta_{f}}}~\text{and}~Z_{\beta_{f}}\coloneqq\sum_{n=1}^N e^{-\beta_{f} E_n}.
\end{equation}
For $m\neq n$, the transition rate $R_{mn}$ can be expressed as
\begin{equation}
R_{mn}=\Gamma e^{-\beta_{f}(B_{mn}-E_n)},
\end{equation}
where $B_{mn}=B_{nm}$ are the barrier coefficients, and $\Gamma$ is a positive constant.

Now, let us formulate the problem.
We consider relaxation that initiates from a thermal state $\Kt{\pi^{i}}$ associated with the inverse temperature $\beta_{i}=(k_{\rm B}T_{i})^{-1}$.
This can be regarded as a temperature quench $T_{i}\to T_{f}$ at time $t=0^{-}$.
Given a pair of cold and hot temperatures, $T_c$ and $T_h$, satisfying $T_{c}<T_{f}<T_{h}$, we investigate the relaxation speed depending on the quench direction, $T_{i}=T_{c} \uparrow T_{f}$ (heating) and $T_{i}=T_{h} \downarrow T_{f}$ (cooling).
The degree of nonequilibrium (or free energy) of each initial state is the same, 
\begin{equation}\label{eq:init.cond}
D({\pi^{c}}\|{\pi^{f}})=D({\pi^{h}}\|{\pi^{f}}),
\end{equation}
where the relative entropy between two distributions $\Kt{p}$ and $\Kt{q}$ is given by
\begin{equation}
D({p}\|{q})\coloneqq \sum_np_n\ln(p_n/q_n).
\end{equation}
Note that the characters, $c$ and $h$, are associated with the initial temperatures; therefore, $c$ and $h$ correspond to the heating and cooling processes, respectively.
For convenience, we define $\mca{D}({p})\coloneqq D({p}\|{\pi^{f}})$.
We aim to answer the question of which quench direction fastens the relaxation.
To this end, we first need to quantify the relaxation speed, which can be evaluated by the distance between the system state and the thermal state.
Analogous to Refs.~\cite{Lu.2017.PNAS,Lapolla.2020.PRL}, the relative entropy is used to measure the distance between states.
In thermal relaxation, the relative entropy is closely related to the free energy and irreversible entropy production \cite{Seifert.2012.RPP} as
\begin{align}
\mca{D}(p_t)&=\beta_f[\mca{F}(p_t)-\mca{F}_f],\\
\Sigma_t&=\mca{D}(p_0)-\mca{D}(p_t),
\end{align}
where $\mca{F}(p_t)=\sum_np_n(t)[E_n+k_{\rm B}T_f\ln p_n(t)]$ is the free energy of the distribution $\Kt{p_t}$, $\mca{F}_i\coloneqq-k_{\rm B}T_i\ln Z_{\beta_i}$, and $\Sigma_t$ is the irreversible entropy production.
Note that $\mca{F}(p_{0^+})\neq\mca{F}_i$ since $\mca{F}(p_{0^+})$ and $\mca{F}_i$ are evaluated with temperatures $T_f$ and $T_i$, respectively.
Let $\Kt{p^i_t}$ be the time-evolution distribution corresponding to the initial state $\Kt{\pi^i}$, then heating is said to be faster (slower) than cooling if
\begin{equation}
\mca{D}({p^c_t})<(>)\mca{D}({p^h_t})
\end{equation}
in long times.
Throughout this study, we compare the relaxation speeds of the two quenches in the long-time regime, not over the entire evolution time. Therefore, it is possible that the heating and cooling curves, $\mca{D}(p_t^c)$ and $\mca{D}(p_t^h)$, may intersect at some point.
In what follows, we explain in detail how to determine the relaxation speed in the long-time regime.

Let $0=\lambda_1>\lambda_2>\dots>\lambda_{N}$ be the eigenvalues of the transition matrix and $\{\Kt{v_n}\}_{n}$ be the set of corresponding eigenvectors,
\begin{equation}
\msf{R}\Kt{v_n}=\lambda_n\Kt{v_n}.
\end{equation}
Notably, all eigenvalues are real numbers since matrix $\msf{R}$ satisfies the detailed balance condition \cite{Schnakenberg.1976.RMP}.
The eigenvectors $\{\Kt{v_n}\}$ form a basis for the space $\mbb{R}^N$ with $\Kt{v_1}=\Kt{\pi^f}$, and $\Kt{v_n}$ is a traceless vector for all $n\ge 2$ (i.e., $\BraKet{1}{v_n}=0$ since $\Br{1}\msf{R}=\Br{0}$).
Here, $\Kt{0}$ ($\Kt{1}$) denotes the $N$-dimensional vector with all zero (one) elements.
Therefore, the initial distribution $\Kt{\pi^{i}}$ can be expressed as a linear combination of $\{\Kt{v_n}\}$ as follows:
\begin{equation}
\Kt{\pi^{i}}=\Kt{\pi^{f}}+\sum_{n=2}^{N}\gamma_n^{i}\Kt{v_n},\label{eq:ini.eigvec.decomp}
\end{equation}
where $\gamma_n^{i}$'s are real numbers.
Consequently, the probability distribution at time $t$ can be analytically written in the following form:
\begin{equation}
\Kt{p^i_t}=\Kt{\pi^{f}}+\sum_{n=2}^{N}\gamma_n^{i} e^{\lambda_n t}\Kt{v_n}.\label{eq:time.eigvec.decomp}
\end{equation}
In the long-time limit, the probability distribution $\Kt{p^i_t}$ can be approximated up to the second-order term as
\begin{equation}
\Kt{p^i_t}\simeq \Kt{\pi^{f}}+\gamma_2^{i} e^{\lambda_2 t}\Kt{v_2}.
\end{equation}
Thus, the relaxation speed can be quantified via the value of $|\gamma_2^{i}|$ \cite{Lu.2017.PNAS}.
Accordingly, heating is faster (slower) than cooling if $|\gamma_2^c|<(>)|\gamma_2^h|$ (see Appendix \ref{app:A} for proof).

A closed form of $\gamma_2^i$ can be obtained analytically \cite{Klich.2019.PRX}.
The transition rate matrix $\msf{R}$ can be transformed to a symmetric matrix $\overline{\msf{R}}=[\overline{R}_{mn}]\in\mbb{R}^{N\times N}$ as follows:
\begin{equation}
\overline{\msf{R}}=\msf{F}^{1/2}\msf{R}\msf{F}^{-1/2},
\end{equation}
where $\msf{F}=[F_{mn}]\in\mbb{R}^{N\times N}$ with $F_{mn}=e^{\beta_{f} E_n}\delta_{mn}$.
The elements of matrix $\overline{\msf{R}}$ can be explicitly written in terms of the elements of matrix $\msf{R}$ as
\begin{equation}
\overline{R}_{mn}=e^{\beta_f(E_m-E_n)/2}R_{mn}.
\end{equation}
Notably, matrix $\overline{\msf{R}}$ has the same eigenvalues as $\msf{R}$, and its eigenvectors $\{\Kt{f_n}\}$ are related to those of $\msf{R}$ as $\Kt{f_n}=\msf{F}^{1/2}\Kt{v_n}$.
Moreover, these eigenvectors are mutually orthogonal, $\BraKet{f_m}{f_n}=\MatBraKet{v_n}{\msf{F}}{v_m}\delta_{mn}$.
Multiplying $\Br{f_2}\msf{F}^{1/2}$ on both sides of Eq.~\eqref{eq:ini.eigvec.decomp}, we can show that $\gamma_2^i$ is proportional to the inner product between the initial distribution $\Kt{\pi^{i}}$ and the vector $\Kt{\overline{f}_2}$, given by
\begin{equation}
\gamma_2^{i}=\frac{\MatBraKet{f_2}{\msf{F}^{1/2}}{\pi^{i}}}{\BraKet{f_2}{f_2}}=\frac{\BraKet{\overline{f}_2}{\pi^{i}}}{\BraKet{f_2}{f_2}},
\end{equation}
where $\Kt{\overline{f}_2}\coloneqq \msf{F}^{1/2}\Kt{f_2}$.
Note that $\BraKet{\overline{f}_2}{\pi^{f}}=0$ since $\Kt{f_2}$ and $\Kt{f_1}=\msf{F}^{1/2}\Kt{\pi^f}$ are orthogonal.
Because the sign of $\gamma_n^{i}$ can be absorbed by changing the eigenvectors $\Kt{v_n}\to -\Kt{v_n}$, hereafter, we assume $\gamma_2^{h}\le 0$, which implies that $\BraKet{\overline{f}_2}{\pi^{h}}\le 0$.

\section{Results}
Given the above setup, we now present our main results on the relaxation asymmetry, including numerical illustrations and proofs.

\subsection{Two-state systems}
First, we consider two-state systems, for which universality regarding the relaxation asymmetry can be achieved.

\begin{result}\label{thm:two.state}
For two-state systems, heating is faster than cooling.
\end{result}

Result \ref{thm:two.state} affirmatively validates that the relaxation asymmetry is universal in two-state systems.
Even for two-state systems, it is highly nontrivial that heating is faster than cooling.
For discrete-state Markovian dynamics, the speed of state transformation is constrained by time-antisymmetric dissipation and time-symmetric frenesy (or dynamical activity) \cite{Shiraishi.2018.PRL,Vo.2020.PRE}.
Relaxation from a thermal state at a higher temperature has a higher dynamical activity; more precisely, the average number of jumps over all the stochastic trajectories in the hot two-state system is always greater than that in the cold two-state system.
Consequently, one may intuitively expect that cooling, which has higher dynamical activity, is faster than heating.
However, the result is counterintuitive, implying that the dynamical activity alone cannot account for this relaxation asymmetry.

It is worth discussing the compatibility of Result \ref{thm:two.state} with the counterexamples in Ref.~\cite{Lapolla.2020.PRL}, in which cooling can be faster than heating for some double-well potentials. 
In the counterexamples therein, there is a time-scale separation: a local equilibration arises prior to the terminal exponential relaxation, which is effectively a two-state Markov jump process.
Although heating occurs faster than cooling in the early stages, the asymmetry is inverted in the later stages.
It has been observed that the breaking of the asymmetry is intimately related to the configuration between the intra-well and inter-well entropies \cite{Lapolla.2020.PRL}.
Therefore, the apparent discrepancy between Result \ref{thm:two.state} and the terminal relaxation trend in the counterexamples is due to the intra-well entropic contribution, which is neglected in the Markov jump dynamics considered here.

\begin{figure}[t]
\centering
\includegraphics[width=1.0\linewidth]{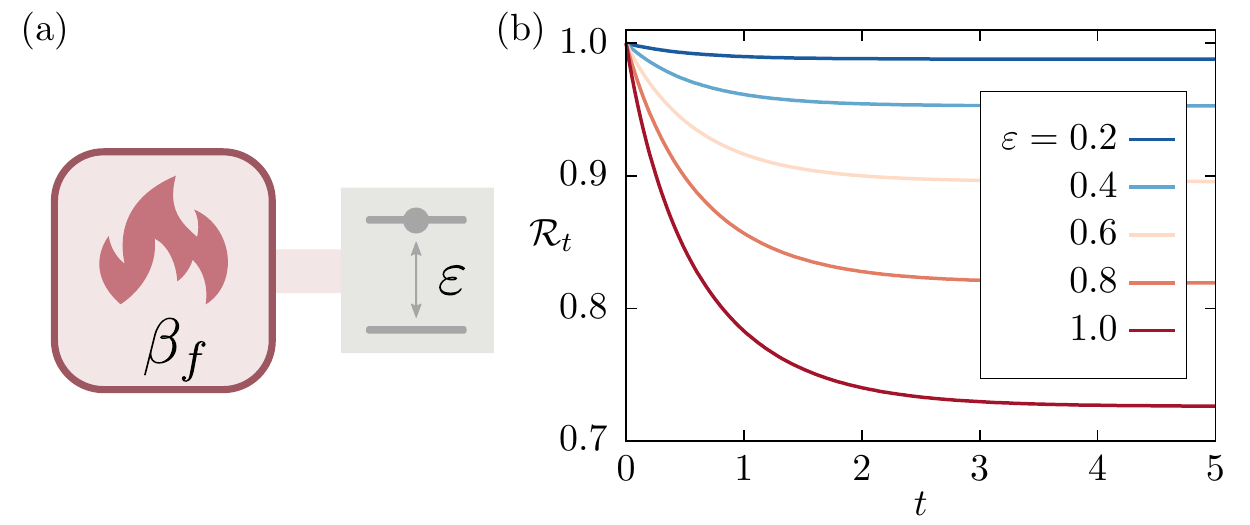}
\protect\caption{(a) Schematic of the two-state system. (b) Numerical illustration of Result \ref{thm:two.state} in the two-state system. The energy gap, $\varepsilon$, is varied from $0.2$ to $1.0$. The solid lines depict ratio $\mca{R}_t$ as a function of time $t$ for each value of $\varepsilon$.}\label{fig:FIG1}
\end{figure}

To illustrate the above result, we use a two-state system [see Fig.~\ref{fig:FIG1}(a)] with the following transition matrix:
\begin{equation}
\msf{R}=\begin{pmatrix}
	-1 & e^{-\beta_f\varepsilon}\\
	1 & -e^{-\beta_f\varepsilon}
\end{pmatrix},
\end{equation}
where $\varepsilon=E_1-E_2>0$ is the energy gap.
We vary the value of $\varepsilon$ while fixing the inverse temperatures as follows: $\beta_f=1$, $\beta_h=0.1$, and $\beta_c$ is uniquely determined via the condition in Eq.~\eqref{eq:init.cond}.
The time variation of ratio $\mca{R}_t\coloneqq\mca{D}(p_t^c)/\mca{D}(p_t^h)$ is plotted as a function of time $t$ in Fig.~\ref{fig:FIG1}(b).
Note that $\mca{R}_t<(>)1$ implies that heating occurs faster (slower) than cooling at time $t$.
As shown, ratio $\mca{R}_t$ is always smaller than $1$, and there is no intersection between the heating and cooling curves at any finite time.
Therefore, it is numerically verified that heating is always faster than cooling.
Although Result \ref{thm:two.state} only indicates the long-time behavior, numerical evidence suggests that heating is faster than cooling for the entire evolution time. 

\begin{proof}[Proof of Result~\ref{thm:two.state}]
It suffices to prove that $|\gamma_2^c|<|\gamma_2^h|=-\gamma_2^h$.
For $N=2$, an arbitrary probability distribution $\Kt{\pi^{i}}$ can be expressed as a point $(x_{i},1-x_{i})$ in the two-dimensional space.
Since $E_1>E_2$, all thermal states, $\Kt{\pi^{c}},\Kt{\pi^{f}}$, and $\Kt{\pi^{h}}$, lie on the segment with $(0,1)$ and $(1/2,1/2)$ as endpoints.
These probability distributions are geometrically illustrated in Fig.~\ref{fig:FIG1}.
In the following, we employ a geometrical approach to prove Result~\ref{thm:two.state}.
From the conditions, $\BraKet{\overline{f}_2}{\pi^{f}}=0$ and $\BraKet{\overline{f}_2}{\pi^{h}}\le 0$, we can conclude that $\BraKet{\overline{f}_2}{\pi^{c}}\ge 0$ or $\gamma_2^{c}\ge 0$.
Thus, it is sufficient to show that
\begin{equation}
\gamma_2^{c}+\gamma_2^{h}<0~\text{or}~\BraKet{\overline{f}_2}{(\pi^{c}+\pi^{h})/2}<0,
\end{equation}
which is equivalent to proving that $(x_{c}+x_{h})/2>x_{f}$.
We can rewrite the equality $D({\pi^{c}}\|{\pi^{f}})=D({\pi^{h}}\|{\pi^{f}})$ as follows:
\begin{equation}
	S({\pi^{h}})-S({\pi^{c}})=(x_{h}-x_{c})\ln\frac{1-x_{f}}{x_{f}},\label{eq:tmp1}
\end{equation}
where $S({p})\coloneqq -\sum_np_n\ln p_n$ is the Shannon entropy of the distribution $\Kt{p}$.
Note that $x_{c}<x_{h}<1/2$ and 
\begin{equation}
g(x_{i})\coloneqq dS({\pi^{i}})/dx_{i}=\ln[(1-x_{i})/x_{i}]
\end{equation}
is a strictly convex function over $x_{i}\in[0,1/2]$ (i.e., the second derivative of $g(x_i)$ with respect to $x_i$ is positive, $g''(x_i)>0$).
Applying the Hermite--Hadamard inequality for $g(x_{i})$, we obtain the following:
\begin{equation}
	\frac{1}{x_{h}-x_{c}}\int_{x_{c}}^{x_{h}}g(x_{i})dx_{i}>g\bra{\frac{x_{c}+x_{h}}{2}},\label{eq:hh.ine}
\end{equation}
or equivalently,
\begin{equation}
	S({\pi^{h}})-S({\pi^{c}})>(x_{h}-x_{c})g\bra{\frac{x_{c}+x_{h}}{2}}.\label{eq:tmp2}
\end{equation}
Combining Eqs.~\eqref{eq:tmp1} and \eqref{eq:tmp2} results in the following inequality:
\begin{equation}
	g(x_{f})>g\bra{\frac{x_{c}+x_{h}}{2}}.
\end{equation}
Since $g(x_{i})$ is a strictly decreasing function, we have $(x_{c}+x_{h})/2>x_{f}$, which completes the proof.
\end{proof}

\begin{figure}[t]
\centering
\includegraphics[width=0.8\linewidth]{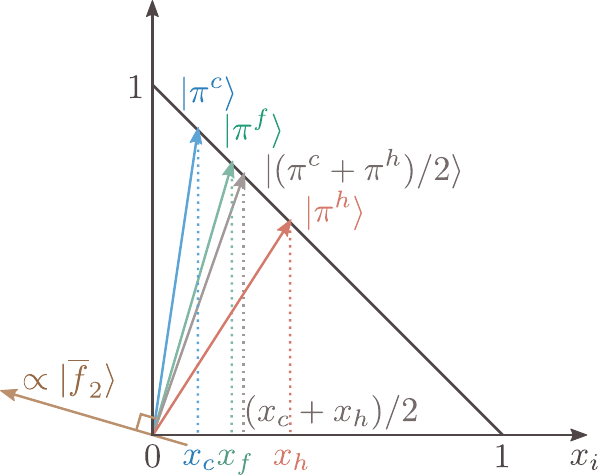}
\protect\caption{Geometrical illustration of probability distributions. Two vectors $\Kt{\overline{f}_2}$ and $\Kt{\pi^{f}}$ are orthogonal, $\BraKet{\overline{f}_2}{\pi^{f}}=0$.}\label{fig:FIG2}
\end{figure}

\subsection{Systems with at least three distinct energy levels}

Next, we consider more general systems that have at least three distinct energy levels; i.e., there exist three indices $1\le i<j<k\le N$ such that $E_i>E_j>E_k$.
For such systems, we obtain the following result.

\begin{result}\label{thm:three.state}
For systems with more than two distinct energy levels, heating can be faster or slower than cooling, depending on the transition rates.
\end{result}

Result \ref{thm:three.state} shows that no universality of the relaxation asymmetry is achieved in the general case.
With an appropriate choice of barrier coefficients, we can construct a discrete-state system with $|\gamma_2^{c}|<|\gamma_2^{h}|$ or $|\gamma_2^{c}|>|\gamma_2^{h}|$ as desired.
This difference between the continuous- and discrete-state systems can be explained as follows.
For simplicity, we consider a single-particle system described by the overdamped Langevin equation.
In this continuous-state system, the particle tends to transit to a close place at any instant time.
By contrast, in discrete-state systems, the particle, in principle, can jump to anywhere, provided the transition rate between these states is positive.
This degree of freedom could lead to a complicated relaxation compared with the continuous-state system.
As shown below, the construction of the transition matrix that determines the magnitude relation between the heating and cooling rates is somewhat artificial; therefore, there may not be a realistic system with such a transition matrix.
We anticipate that the universality of the relaxation asymmetry will be achieved if appropriate constraints are placed on the transition rates.

\begin{figure}[t]
\centering
\includegraphics[width=1.0\linewidth]{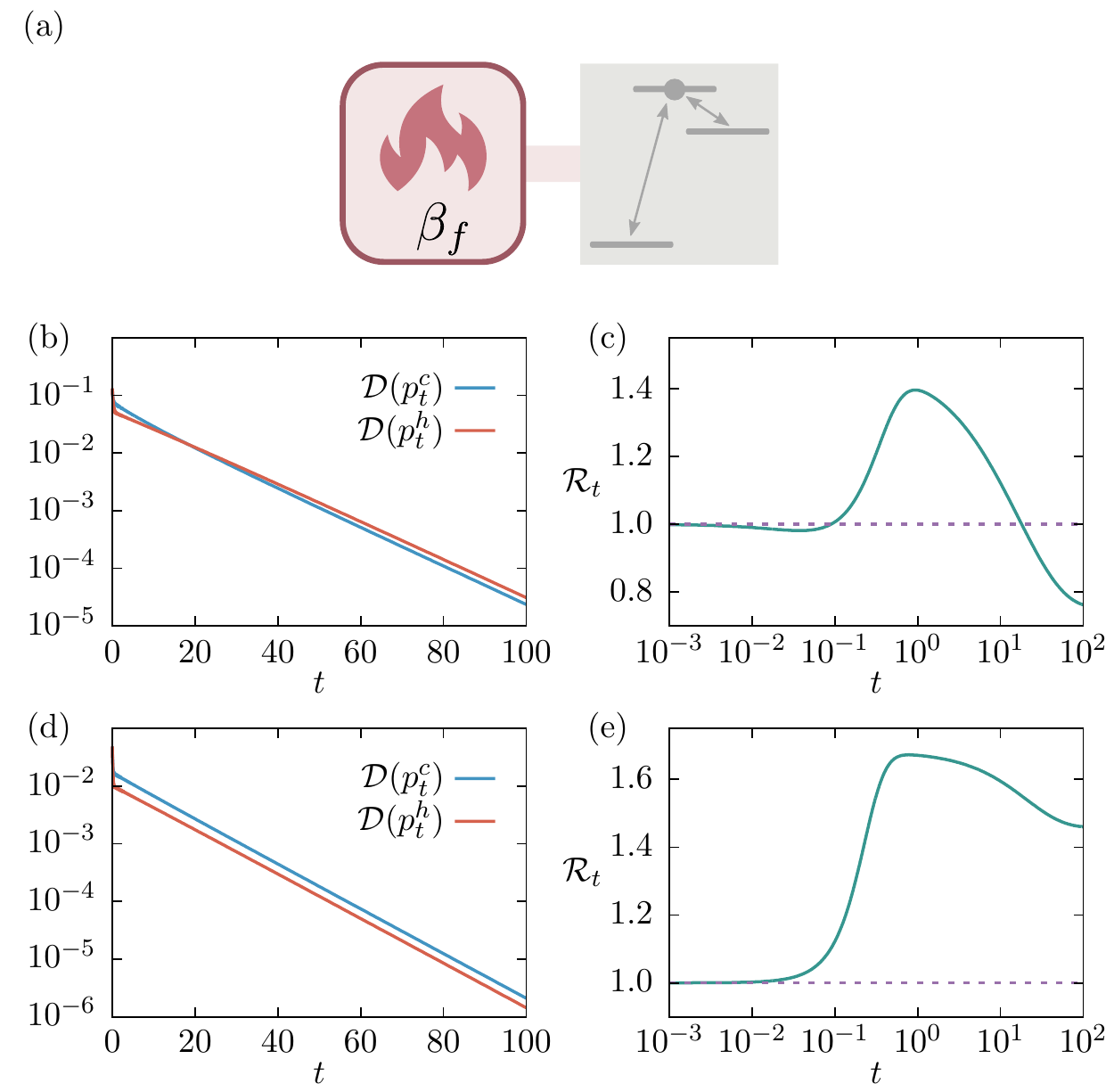}
\protect\caption{(a) Schematic of the three-state system. Numerical illustration of Result \ref{thm:three.state} for the cases of (b)(c) faster heating and (d)(e) faster cooling. (b)(d) The heating and cooling curves, $\mca{D}(p_t^c)$ and $\mca{D}(p_t^h)$, are depicted by blue and red solid lines, respectively. (c)(e) The ratio of $\mca{R}_t=\mca{D}(p_t^c)/\mca{D}(p_t^h)$ for each case is depicted by the green solid line.}\label{fig:FIG3}
\end{figure}

We numerically illustrate Result \ref{thm:three.state} in a three-state system [see Fig.~\ref{fig:FIG3}(a)].
The transition rates are determined using $\Gamma=1$ and
\begin{equation}\label{eq:Bmn}
e^{-\beta_fB_{mn}}=\frac{\gamma_{mn}}{2\cosh[\beta_f(E_n-E_m)]}e^{-\beta_f(E_m+E_n)/2},
\end{equation}
where $\gamma_{12}=\gamma_{21}=0.1$, $\gamma_{13}=\gamma_{31}=10$, and $\gamma_{23}=\gamma_{32}=0$.
The inverse temperatures are fixed as follows: $\beta_f=1$, $\beta_h=0.5$, and $\beta_c$ is uniquely determined via the condition in Eq.~\eqref{eq:init.cond}.
The energy levels are $E_1=2$, $E_2=1.5$, and $E_3=\pm 0.5$.
The numerical results (i.e., the heating and cooling curves and the $\mca{R}_t$ ratio) are plotted in Figs.~\ref{fig:FIG3}(b) and \ref{fig:FIG3}(c) for the $E_3=-0.5$ case and in Figs.~\ref{fig:FIG3}(d) and \ref{fig:FIG3}(e) for the $E_3=0.5$ case.
As shown in Fig.~\ref{fig:FIG3}(c), heating is faster than cooling in the short- and long-time regimes. However, interestingly, two crossing points exist in the intermediate-time regime, where cooling temporarily occurs faster than heating.
While it may be difficult to identify these two crossing points in Fig.~\ref{fig:FIG3}(b), we can clearly see in Fig.~\ref{fig:FIG3}(c) that they appear around times $t=10^{-1}$ and $t=20$.
In contrast, in the $E_3=0.5$ case, Fig.~\ref{fig:FIG3}(e) shows that the $\mca{R}_t$ ratio is always greater than $1$, implying that cooling is faster than heating for the entire evolution time.
These numerical results indicate that the relaxation asymmetry is not universal in three-state systems, which is consistent with the theoretical finding of Result \ref{thm:three.state}.
Here, for simplicity, we have performed numerical calculations with different energy levels in the two cases.
Nevertheless, as shown below, even under the condition where the energy levels are fixed, it is easy to construct a transition matrix that yields the desired relaxation trend.

\begin{proof}[Proof of Result~\ref{thm:three.state}]
We prove the result by analytically constructing a transition rate matrix such that heating is slower than cooling.
A transition rate matrix for the opposite case can also be analogously constructed.
First, one can prove that $\Kt{\pi^{c}},\Kt{\pi^{h}}$, and $\Kt{\pi^{f}}$ are linearly independent (see Appendix \ref{app:B}).
Let $\{\Kt{e_1},\Kt{e_2}\}$ be an orthogonal basis of the space $\mca{S}$ spanned by $\Kt{\pi^{f}}$ and $\Kt{\pi^{h}}$, and $\msf{P}\coloneqq \Mat{e_1}+\Mat{e_2}$ be the projection matrix to the space $\mca{S}$.
Define $\Kt{\overline{f}_2}\coloneqq \Kt{\pi^{c}}-\msf{P}\Kt{\pi^{c}}$, then $\Kt{\overline{f}_2}\neq \Kt{0}$ because $\Kt{\pi^{c}},\Kt{\pi^{h}}$, and $\Kt{\pi^{f}}$ are linearly independent.
Trivially, $\BraKet{\overline{f}_2}{e_1}=\BraKet{\overline{f}_2}{e_2}=0$; thus, $\BraKet{\overline{f}_2}{\pi^{f}}=\BraKet{\overline{f}_2}{\pi^{h}}=0$.
Moreover, $\BraKet{\overline{f}_2}{\pi^{c}}=\BraKet{\overline{f}_2}{\overline{f}_2}>0$ since $\Kt{\overline{f}_2}\neq\Kt{0}$.

Next, we construct a transition rate matrix $\msf{R}$ that results in $|\gamma_2^{c}|>|\gamma_2^{h}|$.
Set $B_{mn}=E_m+E_n$, one can explicitly calculate that the matrix $\overline{\msf{R}}$ has a single zero eigenvalue associated with the eigenvector $\Kt{f_1}=\msf{F}^{1/2}\Kt{\pi^{f}}$, and the remaining eigenvalues are all $-Z_{\beta_{f}}$ \cite{Klich.2019.PRX}.
Let $\mca{U}=\{\Kt{x}\in\mbb{R}^N~|~\BraKet{x}{f_1}=0\}$ be a subspace orthogonal to $\Kt{f_1}$.
Then, there exists an orthogonal basis $\{\Kt{f_2},\Kt{f_3},\dots,\Kt{f_N}\}$ of $\mca{U}$, where $\Kt{f_2}\coloneqq\msf{F}^{-1/2}\Kt{\overline{f}_2}$, since $\BraKet{f_2}{f_1}=0$ (i.e., $\Kt{f_2}\in\mca{U}$).
Obviously, $\Kt{f_n}~(n\ge 2)$ is an eigenvector of $\overline{\msf{R}}$ with the corresponding eigenvalue $-Z_{\beta_{f}}$.
Following the idea in Ref.~\cite{Klich.2019.PRX}, we slightly modify $\overline{\msf{R}}$ as follows:
\begin{equation}
	\overline{\msf{R}}\to\overline{\msf{R}}+\sum_{n=2}^N\epsilon_n\frac{\Mat{f_n}}{\BraKet{f_n}{f_n}}.
\end{equation}
Here, $Z_{\beta_{f}}>\epsilon_2>\dots>\epsilon_N\ge 0$ are small numbers that ensure the positivity of $\overline{R}_{mn}~(m\neq n)$.
It is easy to check that $\overline{\msf{R}}\Kt{f_1}=\Kt{0}$ and $\overline{\msf{R}}\Kt{f_n}=(-Z_{\beta_{f}}+\epsilon_n)\Kt{f_n}$ for all $n\ge 2$.
Now, the matrix $\overline{\msf{R}}$ has $N$ different eigenvalues, and $\Kt{f_2}$ is precisely the eigenvector corresponding to the second-largest eigenvalue $\lambda_2=-Z_{\beta_{f}}+\epsilon_2$.
The transition rate matrix can be recovered as $\msf{R}=\msf{F}^{-1/2}\overline{\msf{R}}\msf{F}^{1/2}$, and the detailed balance condition is satisfied due to the symmetry of $\overline{\msf{R}}$.
With this construction, the relation between $|\gamma_2^{c}|$ and $|\gamma_2^{h}|$ can be clarified as
\begin{equation}
	|\gamma_2^{c}|=\frac{\BraKet{\overline{f}_2}{\pi^{c}}}{\BraKet{f_2}{f_2}}>0=\frac{\BraKet{\overline{f}_2}{\pi^{h}}}{\BraKet{f_2}{f_2}}=|\gamma_2^{h}|,
\end{equation}
which completes the proof.
\end{proof}

\subsection{Degenerate two-level systems}

Last, we consider the remaining case, wherein the energy levels are degenerate to two energy states.
In other words, there exists an index $1\le n<N$ such that $E_1=\dots=E_n>E_{n+1}=\dots=E_N$.
Such degenerated two-level systems are seen in atoms \cite{Margalit.2013.PRA} and have been used to enhance quantum-annealing performance \cite{Watabe.2020.SR} and dissipation-less heat current \cite{Tajima.2021.PRL}.
For convenience, we define the energy gap $\Delta E\coloneqq E_n-E_{n+1}>0$.
Remarkably, we find that, depending on the magnitude of this energy gap, heating can be faster or slower than cooling, regardless of the transition rates.
Details are summarized in the following.

\begin{result}\label{thm:deg.two.state}
If $\beta_{h}\Delta E\ge\ln[n/(N-n)]$, then heating is faster than cooling.
Conversely, if $\beta_{c}\Delta E\le\ln[n/(N-n)]$, then heating is slower than cooling.
\end{result}

Note that Result \ref{thm:two.state} can be derived from Result \ref{thm:deg.two.state} by setting $N=2$ and $n=1$.
Result \ref{thm:deg.two.state} indicates that there are two critical thresholds of the energy gap $\Delta E$.
As the energy gap is above or below these thresholds, a universal conclusion on asymmetry in thermal relaxation can be drawn.
It is also highly nontrivial that the energy gap affects the relaxation speeds of heating and cooling in this way.
When $\Delta E$ is large, the jump from energy state $E_{n+1}$ to $E_n$ is less likely to occur compared with the opposite jump.
Thus, heating is expected to be slower than cooling.
However, counterintuitively, heating is faster than cooling as $\Delta E$ is sufficiently large.
In addition, provided $n\le N/2$, heating is faster than cooling, regardless of the value of $\Delta E$.
This implies that the number of excited states also plays a crucial role in determining the relaxation speed.

\begin{figure}[t]
\centering
\includegraphics[width=1.0\linewidth]{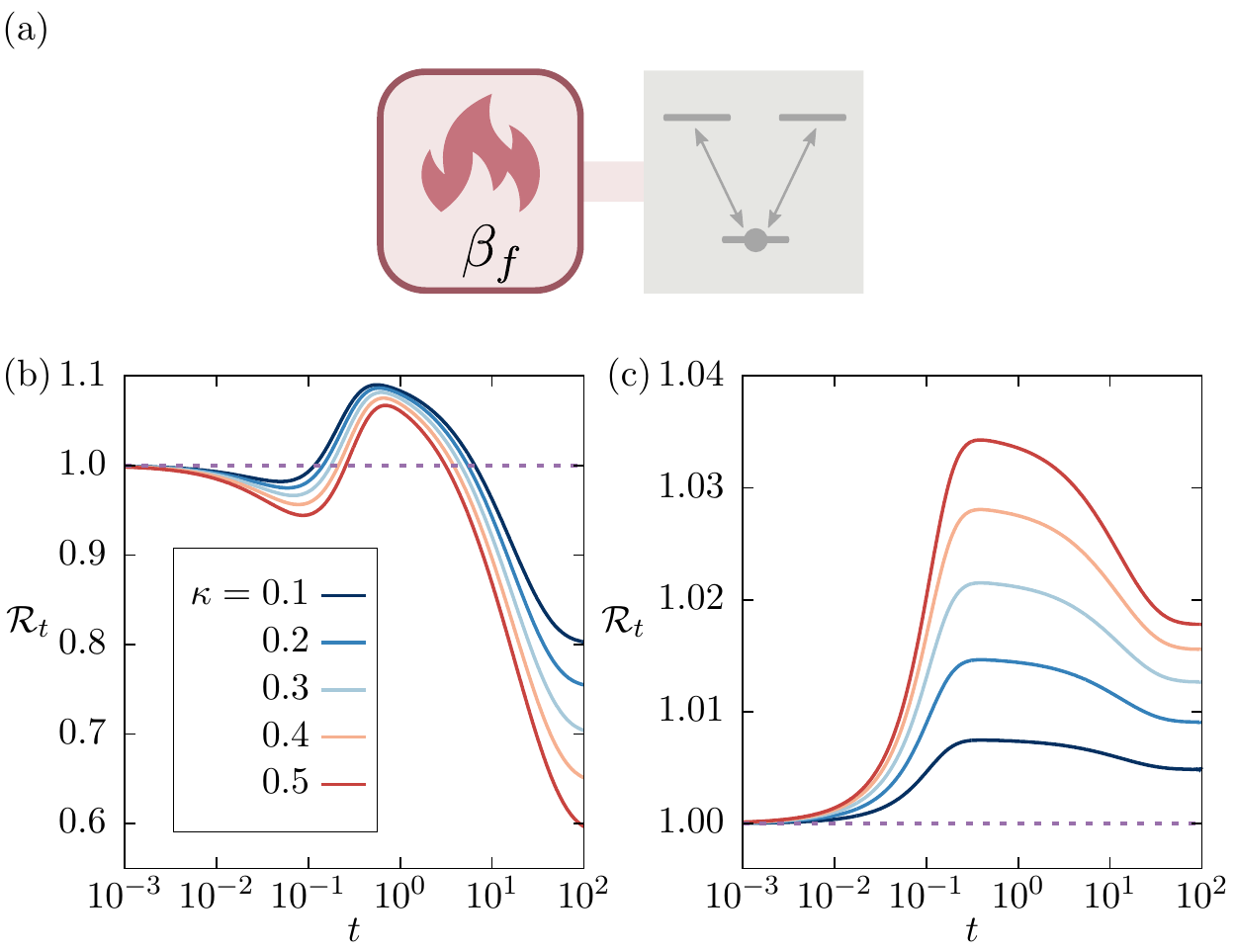}
\protect\caption{(a) Schematic of the three-state system. Numerical illustrations of Result \ref{thm:deg.two.state} for (b) case (i) and (c) case (ii). The solid lines depict the ratio, $\mca{R}_t=\mca{D}(p_t^c)/\mca{D}(p_t^h)$, for each value of $\kappa$. The value of parameter $\kappa$ is varied from $0.1$ to $0.5$.}\label{fig:FIG4}
\end{figure}

Again, we numerically illustrate Result \ref{thm:deg.two.state} in a three-state system with $n=2$ [see Fig.~\ref{fig:FIG4}(a)].
The transition rates are analogously defined as in Eq.~\eqref{eq:Bmn} with $\gamma_{13}=\gamma_{31}=10$, $\gamma_{23}=\gamma_{32}=0.1$, and $\gamma_{12}=\gamma_{21}=0$.
We consider two cases of the parameter setting: (i) $\Delta E=(1+\kappa)\beta_h^{-1}\ln[n/(N-n)]$, $(\beta_f,\beta_h)=(1,0.5)$ and (ii) $\Delta E=\kappa\beta_c^{-1}\ln[n/(N-n)]$, $(\beta_f,\beta_c)=(1,1.5)$.
Here, $\kappa\in(0,1)$ is a tuning parameter.
Note that once the above parameters are provided, the other parameters are uniquely determined.
According to Result \ref{thm:deg.two.state}, cases (i) and (ii) correspond, respectively, to faster heating and faster cooling.
We vary the value of $\kappa$ from $0.1$ to $0.5$ and plot the numerical results of cases (i) and (ii) in Figs.~\ref{fig:FIG4}(b) and \ref{fig:FIG4}(c), respectively.
Figure \ref{fig:FIG4}(b) shows that $\mca{R}_t<1$ in the long-time regime, implying that heating is faster than cooling.
Interestingly, there are crossing points between the heating and cooling curves in the intermediate-time regime.
In both the system considered here and the two-state system considered in the previous section, the energy levels are two-state degenerate. However, the former shows a more complicated relaxation trend than the latter.
For case (ii), conversely, $\mca{R}_t$ is always greater than $1$, implying that cooling always occurs faster than heating. 
Consequently, all these numerical results are consistent with Result \ref{thm:deg.two.state}.

\begin{proof}[Proof of Result~\ref{thm:deg.two.state}]
We employ the same strategy used in proving Result \ref{thm:two.state}.
We prove the former case first, i.e., $\beta_{h}\Delta E\ge\ln[n/(N-n)]$ leads to a faster heating.
It can be observed that all points lying on the segment $\ell$ with $\Kt{\pi^{c}}$ and $\Kt{\pi^{h}}$ as endpoints are thermal states.
It is also evident that $\Kt{\pi^{c}},\Kt{\pi^{f}}$, and $\Kt{\pi^{h}}$ are linearly dependent, i.e., there exists a real number $a\in(0,1)$ such that $\Kt{\pi^{f}}=a\Kt{\pi^{c}}+(1-a)\Kt{\pi^{h}}$.
Since $\BraKet{\overline{f}_2}{\pi^{f}}=0$ and $\BraKet{\overline{f}_2}{\pi^{h}}\le 0$, $\BraKet{\overline{f}_2}{\pi^{c}}\ge 0$ or $\gamma_2^{c}\ge 0$ follows from the continuity of the inner product $\BraKet{\overline{f}_2}{p}$ for $\Kt{p}\in\ell$.
Thus, it suffices to show that $\gamma_2^{c}+\gamma_2^{h}<0$ or $\BraKet{\overline{f}_2}{(\pi^{c}+\pi^{h})/2}< 0$, which is equivalent to proving that $(x_{c}+x_{h})/2>x_{f}$, where $x_{i}=\pi_n^{i}$ completely characterizes the thermal state $\Kt{\pi^{i}}$.
The condition $D({\pi^{c}}\|{\pi^{f}})=D({\pi^{h}}\|{\pi^{f}})$ can be rewritten as follows:
\begin{equation}
	S({\pi^{h}})-S({\pi^{c}})=(x_{h}-x_{c})n\ln\frac{1-nx_{f}}{(N-n)x_{f}}.\label{eq:tmp3}
\end{equation}
Note that $x_{c}<x_{h}\le 1/(2n)$ since $\beta_{h}\Delta E\ge\ln[n/(N-n)]$, and $g(x_{i})\coloneqq dS({\pi^{i}})/dx_{i}=n\ln[(1-nx_{i})/(N-n)x_{i}]$ is a strictly convex function over $x_{i}\in[0,1/(2n)]$.
Applying the Hermite--Hadamard inequality for $g(x_{i})$ and following the same steps as in Eqs.~\eqref{eq:hh.ine} and \eqref{eq:tmp2}, we obtain $(x_{c}+x_{h})/2>x_{f}$, which proves the former case.

When $\beta_{c}\Delta E\le\ln[n/(N-n)]$, one can derive that $1/(2n)\le x_{c}<x_{h}$, and $g(x_{i})$ is a strictly concave function over $x_{i}\in[1/(2n),1]$ (i.e., the second derivative of $g(x_i)$ with respect to $x_i$ is negative, $g''(x_i)<0$).
Applying the Hermite--Hadamard inequality for $g(x_{i})$, we obtain the following:
\begin{equation}
g(x_{f})=\frac{1}{x_{h}-x_{c}}\int_{x_{c}}^{x_{h}}g(x_{i})dx_{i}<g\bra{\frac{x_{c}+x_{h}}{2}},
\end{equation}
or $x_{f}>(x_{c}+x_{h})/2$, which implies $|\gamma_2^{h}|<|\gamma_2^{c}|$.

\end{proof}

\section{Summary and Discussion}
In this study, we elucidated the relaxation asymmetry for discrete-state systems described by Markov jump processes.
We proved that the relaxation asymmetry is universal in two-state systems, but not in generic systems with more than two distinct energy levels.
For systems with two degenerate energy levels, we obtained some universal results indicating that the asymmetry in thermal relaxation depends on the energy gap and the number of excited states.

Notably, the relaxation asymmetry has recently been numerically studied for few-level open quantum systems described by the Lindblad master equations \cite{Manikandan.2021.arxiv}. When the initial density matrix contains no coherence, the quantum systems can be described by classical Markov jump processes with the population distributions. It has been shown that heating is always faster than cooling for two-level systems, whereas heating can be faster or slower than cooling for three- and four-level systems. These numerical demonstrations are consistent with Results \ref{thm:two.state} and \ref{thm:three.state}, thus affirmatively supporting our theoretical findings.

Although the relaxation asymmetry is not universal in generic discrete-state systems, universality may be achieved by imposing some constraints on the transition rates.
Such a question requires further investigation and will be addressed in future work.
It would also be interesting to study the relaxation asymmetry for non-Markovian systems \cite{Yang.2020.PRE}.
For instance, the relaxation of some non-Markovian processes, such as a tagged particle in a single file and the end-to-end distance in a Rouse polymer, has been studied in Ref.~\cite{Lapolla.2020.PRL}.
Another possible direction involves formulating and investigating the relaxation asymmetry in open quantum systems \cite{Breuer.2002,Nava.2019.PRB} where coherence is present in the initial state.

\begin{acknowledgements}
We thank Keiji Saito for insightful discussions.
This work was supported by Ministry of Education, Culture, Sports, Science and Technology (MEXT) KAKENHI Grant No. JP19K12153.
\end{acknowledgements}

\appendix

\section{Quantification of relaxation speed}\label{app:A}
Given $\Kt{\pi^{i}}=\Kt{\pi^{f}}+\sum_{n=2}^{N}\gamma_n^{i}\Kt{v_n}$ and $|\gamma_2^{c}|<|\gamma_2^{h}|$, we will show that $\mca{D}({p^{c}_t})<\mca{D}({p^{h}_t})$ as $t\to\infty$.
Note that $\Kt{p^{i}_t}=\Kt{\pi^{f}}+\sum_{n=2}^{N}\gamma_n^{i} e^{\lambda_n t}\Kt{v_n}$.
In the long-time limit, the term $\sum_{n=2}^{N}\gamma_n^{i} e^{\lambda_n t}\Kt{v_n}$ vanishes.
Since
\begin{equation}
D({p}\|{p+dp})=\sum_{k=1}^N\frac{dp_k^2}{p_k}+O(\Delta^3),
\end{equation}
where $\Delta=\sum_{k=1}^N|dp_k|$, we can approximate
\begin{equation}
\mca{D}({p^{i}_t})=\sum_{k=1}^N\frac{\bra{\sum_{n=2}^{N}\gamma_n^{i} e^{\lambda_n t}v_{nk}}^2}{\pi_k^{f}}+O(e^{3\lambda_2t}).
\end{equation}
Here, $\Kt{v_n}=[v_{n1},\dots,v_{nN}]^\top$.
Consequently, we have
\begin{equation}
\begin{aligned}[b]
\mca{D}({p^{h}_t})-\mca{D}({p^{c}_t})&=\bras{(\gamma_2^{h})^2-(\gamma_2^{c})^2}\sum_{k=1}^N\frac{v_{2k}^2}{\pi_k^{f}}e^{2\lambda_2t}\\
&+\sum_{\substack{m,n\ge 2\\\max\{m,n\}>2}}a_{mn}e^{(\lambda_m+\lambda_n)t}+O(e^{3\lambda_2t}),
\end{aligned}
\end{equation}
where $a_{mn}$ are constants.
The first term on the right-hand side is positive since $|\gamma_2^{c}|<|\gamma_2^{h}|$.
The remaining terms may be negative; however, they are negligible compared with the first term in the long-time limit.
Therefore, $\mca{D}({p^{c}_t})<\mca{D}({p^{h}_t})$ as $t\to\infty$.

\section{Proof of linear independence}\label{app:B}
To prove the linear independence of $\Kt{\pi^{c}},\Kt{\pi^{f}}$, and $\Kt{\pi^{h}}$, it is sufficient to show that the determinant of the following matrix is negative.
\begin{equation}
\msf{X}=\begin{pmatrix}
	e^{-\beta_{c}E_1} & e^{-\beta_{c}E_2} & e^{-\beta_{c}E_3}\\
	e^{-\beta_{f}E_1} & e^{-\beta_{f}E_2} & e^{-\beta_{f}E_3}\\
	e^{-\beta_{h}E_1} & e^{-\beta_{h}E_2} & e^{-\beta_{h}E_3}
\end{pmatrix}.
\end{equation}
Here, $E_1>E_2>E_3$ and $\beta_{c}>\beta_{f}>\beta_{h}$.
Without loss of generality, we assume that $E_3=0$ and $\beta_{h}=0$.
In this case, the determinant can be calculated as follows:
\begin{equation}
|\msf{X}|=(1-e^{-\beta_{c}E_1})(1-e^{-\beta_{f}E_2})-(1-e^{-\beta_{c}E_2})(1-e^{-\beta_{f}E_1}).
\end{equation}
Therefore, $|\msf{X}|<0$ is equivalent to
\begin{equation}
\frac{1-e^{-\beta_{c}E_1}}{1-e^{-\beta_{f}E_1}}<\frac{1-e^{-\beta_{c}E_2}}{1-e^{-\beta_{f}E_2}}.\label{eq:app.tmp1}
\end{equation}
Set $h(x)\coloneqq (1-e^{-\beta_{c}x})/(1-e^{-\beta_{f}x})$, Eq.~\eqref{eq:app.tmp1} is equivalent to $h(E_1)<h(E_2)$.
We need only prove that $h(x)$ is a strictly decreasing function over $x>0$.
Taking the derivative of $h(x)$ with respect to $x$, we have
\begin{equation}
\frac{dh(x)}{dx}=\frac{e^{(\beta_{f}-\beta_{c})x}[\beta_{c}(e^{\beta_{f}x}-1)-\beta_{f}(e^{\beta_{c}x}-1)]}{(e^{\beta_{f}x}-1)^2}.
\end{equation}
Since
\begin{align*}
\beta_{c}(e^{\beta_{f}x}-1)-\beta_{f}(e^{\beta_{c}x}-1)&=\sum_{n=1}^{\infty}\frac{x^n}{n!}(\beta_{c}\beta_{f}^n-\beta_{f}\beta_{c}^n)\\
&=\sum_{n=1}^{\infty}\frac{x^n}{n!}\beta_{c}\beta_{f}(\beta_{f}^{n-1}-\beta_{c}^{n-1})\\
&<0~(\because~0<\beta_{f}<\beta_{c}),
\end{align*}
we have $dh(x)/dx<0$, which completes the proof.

\end{document}